\documentclass[a4paper]{article}

\usepackage{INTERSPEECH2021}
\usepackage{csvsimple}
\usepackage[table,xcdraw]{xcolor}
\usepackage{pgfplots}
\usepackage{hyperref}
\hypersetup{
    colorlinks=true,
    linkcolor=blue,
    filecolor=magenta,      
    urlcolor=cyan,
}
\usepackage{subcaption}
\usepackage{adjustbox}
\usepackage{float}
\newcommand{\cmmnt}[1]{}

\usepackage{tikz}

\title{Conditional Independence for Pretext Task Selection \\ in Self-Supervised Speech Representation Learning }
\name{Salah Zaiem$^{1,2}$, Titouan Parcollet$^2$, Slim Essid$^1$}
\address{
  $^1$LTCI, Télécom Paris, Institut Polytechnique de Paris, Palaiseau, France\\
  $^2$Avignon Université, LIA, Avignon, France}
\email{zaiemsalah@gmail.com}

\newcommand{\ie}{\textit{i.e. }}

\begin{document}

\maketitle

\begin{abstract}

Through solving pretext tasks, self-supervised learning (SSL) leverages unlabeled data to extract useful latent representations replacing traditional input features in the downstream task. A common pretext task consists in pretraining a SSL model on pseudo-labels derived from the original signal. This technique is particularly relevant for speech data where various meaningful signal processing features may serve as pseudo-labels. However, the process of selecting pseudo-labels, for speech or other types of data, remains mostly unexplored and currently relies on observing the results on the final downstream task. Nevertheless, this methodology is not sustainable at scale due to substantial computational (hence carbon) costs. Thus, this paper introduces a practical and theoretical framework to select relevant pseudo-labels with respect to a given downstream task. More precisely, we propose a functional estimator of the pseudo-label utility grounded in the conditional independence theory, which does not require any training. The experiments conducted on speaker recognition and automatic speech recognition validate our estimator, showing a significant correlation between the performance observed on the downstream task and the utility estimates obtained with our approach, facilitating the prospection of relevant pseudo-labels for self-supervised speech representation learning.

\end{abstract}

\section{Introduction}
 Self-supervised learning (SSL) methods usually solve pretext tasks to learn useful representations, taking advantage of the available unlabeled data, whether it is text, images\cite{doersch2016unsupervised} or audio samples \cite{Arandjelovic_2018_ECCV}, for better performance on downstream tasks. Thus, this approach helps improving the results obtained on the considered task without relying on costly and sometimes imprecise manual annotations.
 
 For instance, SSL models have recently been proposed to benefit from large amounts of unlabeled speech data, leading to state-of-the-art results in various speech processing tasks such as automatic speech recognition (ASR) or speech enhancement \cite{wang2020selfsupervised}. Various paradigms have thus been introduced including: predictive coding \cite{baevski2020wav2vec,Liu_2020,song2020speechxlnet,Zhang2020}, pseudo-label learning \cite{pascual2019learning,ravanelli2020multitask}, auto-encoding techniques \cite{Renshaw2015ACO,algayres:hal-02977539}, generative modelling \cite{khurana2020convolutional} or contrastive learning \cite{Saeed2020,jiang2020speech}. 
 
  Pretext tasks may be defined through a choice of pretext labels, hereafter referred to as \textit{pseudo-labels}. The automatic generation of pseudo-labels is a common technique to conceive SSL models in many application domains such as computer vision \cite{noroozi2017unsupervised}, music processing \cite{hung2019multitask} and speech processing \cite{pascual2019learning, shukla}. In the latter scenario, examples of pseudo-labels include, but are not limited to, pitch estimators, energy-based features, voicing state... As a matter of fact, decades of research in signal processing offer a wide range of potential features to be considered as pseudo-labels.
  
 However, the process of selecting the most relevant signal features among the ones present in the speech processing literature is still essentially driven by intuition or empirical validation. Empirical assessment implies a heavy computational load due to a large number of required pretraining and fine-tuning steps. This results in a substantial carbon footprint and may lead to intractability issues. In this work, we aim to provide a clear procedure for a theoretically motivated and efficient pseudo-label selection. This is achieved by introducing a function that estimates the utility of considering a given pseudo-label. 
 

Despite few recent works on the theory of contrastive learning \cite{chen2020simple, oord2018representation, lee2020predicting, Arora2019} the literature on the theoretical foundations of pseudo-label-based SSL remains extremely scarce. Lee and al.\cite{lee2020predicting} proposed a novel approach building a link between the downstream-task performance and the conditional independence (CI) between a pseudo-label and the training samples given the downstream labels. However, their experiments are not related to speech and are restricted to pseudo-labels with an enforced strict conditional independence which is not the case of traditional speech features. On the other hand, numerous pseudo-labels have been empirically tested to generate useful latent speech representations \cite{pascual2019learning}. Pascual and al.\cite{pascual2019learning} introduced a novel SSL method for speech referred to as \textit{PASE} alongside with a thorough empirical ablation study on the considered pseudo-labels highlighting the most influential ones. A similar study has been done on music data for instrument recognition by Hung and al.\cite{hung2019multitask}. Nevertheless, neither works provide a prior quantitative motivation to justify the pseudo-labels selection that was thus potentially performed with grid or random searches. In short, and to the best of our knowledge, explaining and motivating the selection process of pseudo-labels remain open research questions for SSL on speech data.
Therefore, the main contributions of our work are threefold:
\begin{enumerate}
    \item Propose a method to compute an estimate of the conditional independence between the pretext task and the downstream speech samples given the downstream label.
    \item Show that this estimate predicts well the utility of a given pseudo-label for a given downstream task, as it correlates highly with the downstream performance on two tasks: ASR (TIMIT) and speaker recognition (VoxCeleb).
    \item Release the code base developed with SpeechBrain \cite{speechbrain} for replication and to encourage further investigations.\footnote{\url{https://github.com/salah-zaiem/Pseudo-Label-Selection}}
\end{enumerate}
The conducted experiments demonstrate that the proposed method allows a more intelligent, \ie better informed, pretext task selection in self-supervised learning settings.

\section{Conditional Independence Estimation}
This section details the computation of the conditional independence estimate that we propose as a candidate for the measure of a pseudo-label utility. First, we motivate this choice with a precise description of the theoretical background. Then, we describe the computation steps. 

 \begin{figure*}[t!]
  \centering
  \includegraphics[width=0.8\linewidth, scale=0.2]{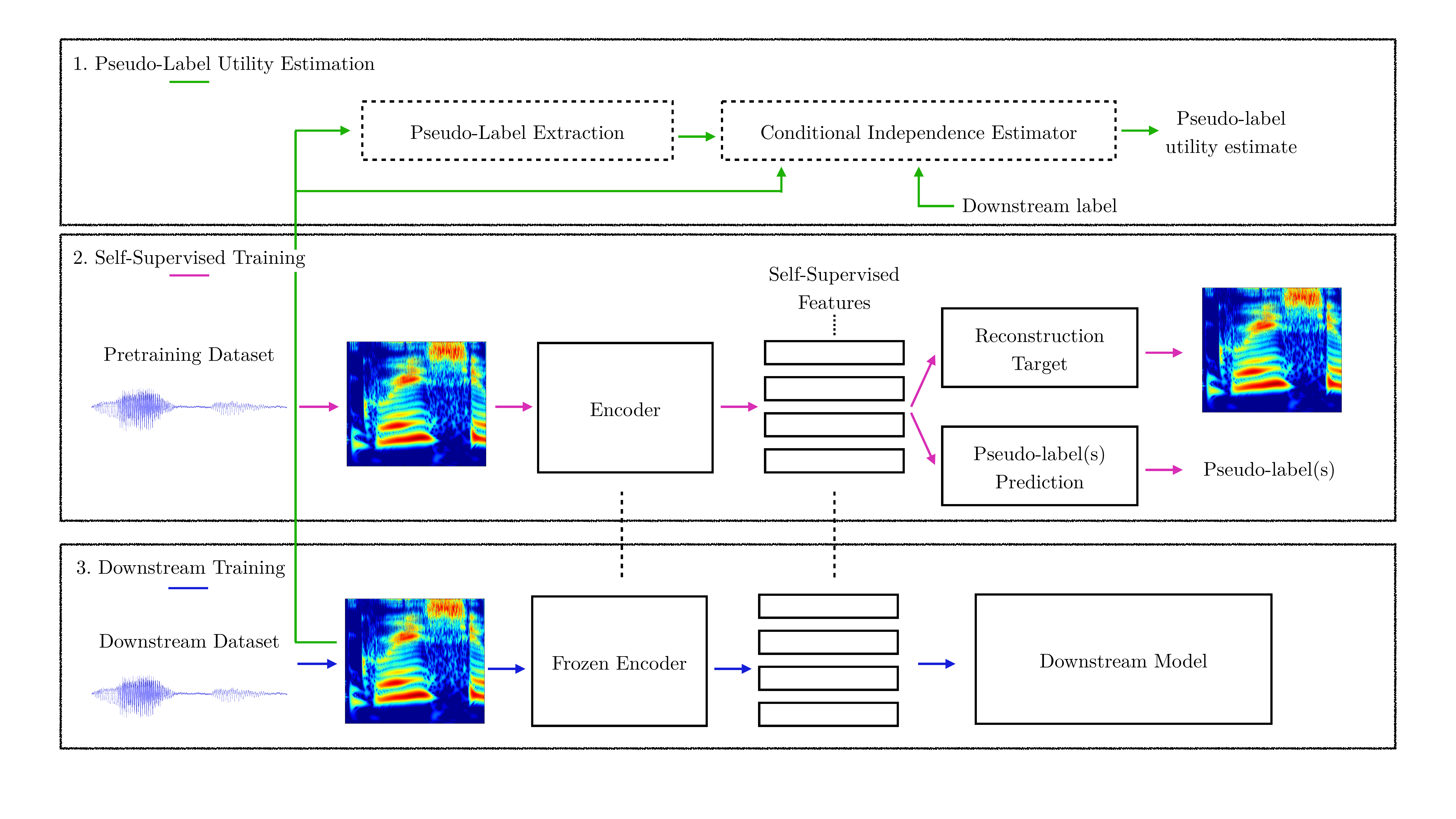}
  \caption{Illustration of the entire training pipeline including estimation, SSL and the downstream parts. The three steps are depicted: 1. estimate the pseudo-label utility; 2. SSL training with the candidate pseudo-label; 3. Train on the downstream task with the pretrained SSL model. The candidate pseudo-label is selected among various candidates based on its conditional independence score. }
  \label{diagram}
\end{figure*}

 Let $X$, $Y$ and $Z$ be, respectively, the downstream data points, the downstream labels and the pseudo-labels which we decide to learn to predict. Let also $\mathcal{C}$ be the set of possible downstream classes. As an example, if we consider speaker recognition as a downstream task, $X$ would be the speech samples, $Y$ the speaker IDs, $\mathcal{C}$ the set of unique speaker IDs, and $Z$ a generated signal feature, such as the fundamental frequency.  Let  $X =(x_i)_{i \in \{0, \ldots, M\}}$ with $M$ being the cardinal of $X$. Each $x_i$ is a speech sample, represented as a Mel band spectrogram. Every sample  $x_i$ has a corresponding downstream label $y_i$ and an automatically generated pseudo-label $z_i$. In the considered cases, $y_i$ is always discrete, whether it is the speaker ID for speaker recognition or the phone for ASR. To every $x_i$, corresponds one value $z_i$, which is the mean of the framewise pseudo-label values.

 As stated above, Lee and al.\cite{lee2020predicting} linked the utility of a pseudo-label ($Z$) to the conditional independence between $Z$ and $X$ given $Y$. In other terms, given the labels $Y$, we want to \textit{quantify how much we can possibly predict the pseudo-labels $Z$ without knowing much about $X$}. In this work, the authors demonstrated that under certain assumptions, the downstream classifier error was bounded by a function of the downstream training set size, and a measure of the conditional dependence. More precisely, the main theorem shows that the bounding function decreases linearly with the downstream-task dataset size ($M$) and quadratically with the conditional independence, thus making conditional independence a potential good estimator of pseudo-label utility.
 The principal issue with conditional independence is the difficulty of computing good estimates of this quantity on realistic data. For our measure, we choose to rely on a kernel-based independence criterion: the Hilbert Schmidt Independence Criterion (HSIC) \cite{gretton}. HSIC has already been proven successful for textual data in testing statistical dependence between translated sentences\cite{gretton}. Our choice is motivated by the fact that kernel-based techniques facilitate handling multivariate and complex data, as the estimation then boils down to the computation of a similarity measure between speech samples. 

Here are the steps to compute our CI estimate of a pseudo-label $Z$ for a downstream task $(X,Y)$, inspired by \cite{gretton}, with further details below: 
\begin{enumerate}
    \item Regroup the samples $X$ by the downstream classes $\mathcal{C}$.
    \item Embed the speech samples $X$ into fixed-size representations.
    \item Compute for every downstream class $c \in \mathcal{C}$, the kernel matrices $K_c$ and $L_c$ containing the similarity measures for the speech samples, and the pseudo-labels, respectively.
    \item Compute the independence test for every split group using $K_c$ and $L_c$, and aggregate the estimations.
\end{enumerate}

We start by splitting the speech samples according to the downstream classes. To obtain the similarity matrices, the second step aims to compute fixed-size embeddings for the speech samples. We wanted to avoid any training for this phase, so we chose the gaussian downsampling method \cite{holzenberger:hal-01888708} detailed thereafter. After the Mel spectrogram extraction, a speech sample becomes a sequence of $L$ input feature vectors of dimension $D$. The goal is, for varying $L$, to obtain fixed size embeddings of size $N \times D$, with $N$ a fixed hyper-parameter for all the samples. To do so, the sequence is divided into $N$ parts. In each part, we compute a Gaussian average of the input frames around the center of the considered part with the standard deviation $\sigma_{gd}$ being another hyper-parameter. This leads for any sample to a $N \times D$ tensor without any training procedure.

Therefore for two speech samples $x_i$ and $x_j$, holding two pseudo-label values $z_i$ and $z_j$, the coefficients of our kernel similarity matrices are:
\begin{equation}
\begin{split}
K_{ij} &= K(x_i,x_j) = cos(GD(x_i) ,GD(x_j) ),   \\
L_{ij} &= RBF(z_i, z_j), 
\end{split}
\end{equation}

with $GD(.)$ the Gaussian Downsampling function, $cos(.,.)$ the cosine similarity, and $RBF(., .)$ the Radial Basis Function kernel defined as: 
\begin{equation}
\begin{split}
    cos(x,x') &= \frac{trace(x^T x')}{||x|| . ||x'||}, \\
    RBF(x,x') &= exp (-\frac{|| x-x'||^2}{2 \sigma^2}),
\end{split}
\end{equation}
with $\sigma$ being the width of the RBF kernel and $trace(.)$ being the sum of elements on the main diagonal.

For each group of samples sharing the same downstream class $c \in C$, we compute the matrices $K_c$ and $L_c$. $K_c$ and $L_c$ correspond to the definitions above, but restricted to the points with $c$ as a downstream label.
For each downstream class $c$, and as in \cite{gretton} the HSIC value is:
\begin{equation}
HSIC_c(X, Z) = \frac{1}{n_c^2} trace(K_c H_c L_c H_c), 
\end{equation}
with $H_c= I_{n_c} - \frac{1}{n_c}1_{n_c} 1_{n_c}^T $, $n_c$ being the number of points with downstream label $c$ and $1_{n_c}$ a vector of ones of size $n_c \times 1$.

The HSIC value is used to characterise the independence of two variables. This value corresponds to the Hilbert norm of their cross-covariance matrix. Intuitively, the HSIC value is high if samples similar in $K$ are similar in $L$. Therefore, the lower this value is, the more independent the two arguments of HSIC are. We enforce the condition on $Y$ by splitting by groups of points sharing the same downstream label.

The final value for a given pseudo label and a downstream task is a weighted mean taking into account the number of samples per downstream class. So with $M$ being the total number of points, and $n_c$ being the number of points having $c$ as their downstream label:
\begin{equation}
HSIC(X,Z |Y) = \frac{1}{M}\sum_{c \in \mathcal{C}} HSIC_c(X, Z) \times n_c.
\end{equation}

\section{Datasets and Experimental Setup}
This sections details the experiments validating the CI measure described above. The estimator is evaluated on two speech tasks that involve different aspects of the audio signal: automatic speech recognition (TIMIT) and speaker recognition (VoxCeleb). Thus, three different datasets are used in this work, one per downstream task considered and a common one for the self-supervised pretraining (Common Voice). CI is computed on both tasks for a list of pseudo-labels, mainly related to prosody and aggregates of spectral descriptors, given in Table \ref{tab:word_styles}. These features are extracted using the OpenSmile library \cite{opensmile}. They have been chosen among the features described in the feature selection literature for various speech tasks.

\begin{table}[t]
  \caption{Candidate speech pseudo-labels and descriptions. }
  \label{tab:word_styles}
  \centering
  \begin{tabular}{ll}
    \toprule
    \textbf{Feature}      & \textbf{Description}                \\
    \midrule
    Loudness                    & Intensity \& approx. loudness                                       \\
    F0                   & Fundamental Frequency                                 \\
    Voicing              & Voicing Decision                     \\
    Alpha Ratio  \cite{alpha}                 & Ratio of spectrum intensity \% 1000 Hz                               \\

    Zero Crossing Rate                & Zero crossing number per frame                 \\
    RastaSpec L1Norm              & L1 Norm of Rasta Spectrum  \cite{rasta}                                    \\
    log HNR  \cite{hnr}   & log of Harmonicity to Noise Ratio        \\
    \bottomrule
  \end{tabular}
\end{table}

\subsection{Datasets}
The train set of the English Common Voice dataset (version $6.1$) \cite{ardila2020common} is used for SSL pretraining ($900$ hours). Common Voice is a collection of speech utterances from worldwide users recording themselves from their own devices. Hence, the closeness to natural settings makes it a suitable choice for self-supervised learning. We remove from Common Voice the sentences lasting more than $10$ seconds, as they often contain long silence parts due to open microphones. 

VoxCeleb1 \cite{Nagrani_2017} is used for the speaker recognition task. The training set contains $148,642$ utterances from $1251$ different speakers. To compute the conditional independence estimates, we restricted ourselves, for tractability issues, to the utterances of $50$ different speakers (the detailed list is given in the released repository\footnote{\label{note1}\url{https://github.com/salah-zaiem/Pseudo-Label-Selection}}). 

TIMIT \cite{timit} is considered for the ASR task. It is composed of a standard $462$-speakers
training set, a $50$-speakers development set and a core test set of $192$ sentences for a total of $5$ hours of clean speech. For the CI estimation, and to get discrete labels to split on, we cut the sentences at the phone level, using the official transcripts. 

\subsection{Self-supervised training}
Based on previous work conclusions \cite{ravanelli2020multitask, jiang2020speech}, apart from the pseudo-label to be tested, our self-supervised model learns to reconstruct the input Mel spectrograms, and to compute $40$-dimensioned MFCC feature vectors. These targets are kept to avoid information loss harming heavily downstream performances. Inspired by the PASE model \cite{ravanelli2020multitask, pascual2019learning}, the model consists of an encoder followed by small predictors limited in capacity. 
Our pretraining model takes as input the speech samples as $80$-Mel band spectrograms. The frame size is $25$ms and hop size $10$ms. The encoder outputs the same number of frames each corresponding to a $256$-dimensional feature embedding. These new embeddings are the ones that will be subsequently extracted for the downstream-task retraining. The new features are then fed to the reconstruction workers and to the pseudo-label prediction. To facilitate the learning, pseudo-label are predicted at the frame level. Predictions are made on top of the encoder with a single linear layer with a PReLu \cite{he2015delving} activation. The final loss is the sum of every predictor' loss: MSE loss for the reconstructions, and $\ell_1$-loss for the considered pseudo-label. The encoder is composed of three distinct parts: a VGG-like features extractor, a bidirectional LSTM, and a two-layered dense neural network with leakyRelu activations. The AdaDelta optimizer is used to update the weights with $1$ as a starting learning rate, $\rho=0.8$ and $\epsilon=10^{-8}$. For every pseudo-label, the network is trained for $10$ epochs. For the CI estimator, as in the work presenting the gaussian downsampling method\cite{holzenberger:hal-01888708}, we fix $N=20$ and $\sigma_{gb} =0.07$. After a few trials aiming to get spaced similarity measures, we fixed the RBF kernel width to $\sigma=0.05$. All the architectures details and hyperparameters can be found in the repository\textsuperscript{\ref{note1}}.

\begin{table}[h]
  \caption{EER/PER values when learning to predict multiple pseudo-labels jointly. "Best" corresponds to the selection of the pseudo-labels with low CI estimator, "Worst" for the high ones. EER is shown for VoxCeleb experiments and PER for TIMIT. The middle column shows the selected pseudo-labels in the experiment.}
  \label{tab:regrouping}
  \centering
  \begin{tabular}{lll}
    \toprule
    \textbf{Experiment}      & \textbf{Pseudo Labels}  & \textbf{EER/PER}              \\
    \midrule
    Best VC  & F0 /log HNR / AlphaR             & 6.40                                        \\
Worst VC           & Loud/ZCR/RastaL1/ Voicing            &  7.33 

\\

Best TIM & F0/RastaL1/AlphaR/log HNR  &  15.35

\\
Worst TIM & Voicing/ Loud/ ZCR & 16.77
\\
    \bottomrule
  \end{tabular}
\end{table}
    
\subsection{Downstream Training}
After extracting the Mel spectrograms from the downstream training data, these are fed to the frozen SSL pretrained encoder to get the self-supervised features. For the ASR retraining, we considered a speech recognition model based on CTC and attention from the SpeechBrain \cite{speechbrain} library. The encoder is similar to the self-supervised training one. It is combined with a location-aware attentive recurrent (LiGRU) decoder \cite{Ravanelli_2018} jointly trained with the CTC loss \cite{kim2017joint}. The model is trained for $50$ epochs on the official train, dev, and test TIMIT sets. Performance is reported in term of Phone Error Rate (PER).

For VoxCeleb, we trained an XVector model\cite{snyder} for $10$ epochs with the frozen SSL features as input. The training recipe follows the one released within SpeechBrain \cite{speechbrain}. The extracted speaker embeddings are tested on the enrol and test splits using PLDA \cite{plda} as a similarity metric. Performance is reported in term of Equal Error Rate (EER)

We chose not to use any data augmentation or added noise during the training to avoid possible interference in our analysis. As a little variance was observed when changing the random seeds used for the TIMIT runs ($\Bar{\sigma} =0.20$), the results presented are the mean of three different runs from three different seeds.

\section{Results}

\begin{figure}[h] 
\centering
\begin{adjustbox}{width=\linewidth}

\begin{tikzpicture}
\pgfplotsset{set layers}

\begin{axis}[
    width=\linewidth, height=7cm, 
    scale only axis,
    grid = major,
    grid style={dashed, gray!30},
    xmin=0.5,   
    xmax=7.5,  
    ymin=-0.25,   
    ymax=4.5,  
    axis background/.style={fill=white},
    axis y line*=right,
    tick align=outside,
     xticklabels={F0,Voicing,logHNR,specRastaL1, Loudness,PCM ZCR,Alpharatio},xtick={1,2,3,4,5,6,7},
  x tick label style={rotate=60,anchor=east}]
    mark repeat={600},
    ylabel near ticks,
    xlabel near ticks,
    xlabel style={text width=5cm},
    xlabel style={align=center},
    xlabel={test},
    legend pos=north west,
    legend columns=1,
    legend style={
            /tikz/column 2/.style={
                column sep=2pt,
            }
    }
    ]
\addplot+[mark=o, color=purple,line width=1pt] file {cit.txt};
\node [above=4 , text=black] at (axis cs:  1,  0.21) {$0.21$};
\node [below=5, text=black] at (axis cs:  2,  0.71) {$0.71$};
\node [above=4, text=black] at (axis cs:  3,  0.17) {$0.17$};
\node [above=4, text=black] at (axis cs:  4,  0.43) {$0.43$};
\node [above=4, text=black] at (axis cs:  5,  0.85) {$0.85$};
\node [above=4, text=black] at (axis cs:  6,  0.80) {$0.80$};
\node [above=5, text=black] at (axis cs:  7,  0.07) {$0.07$};

\addlegendentry{CI TIMIT}
\end{axis}
\begin{axis}[
    width=\linewidth, height=7cm, 
    scale only axis,
    grid = major,
    grid style={dashed, gray!30},
    xmin=0.5,   
    xmax=7.5,  
    ymin=15,   
    ymax=20,  
    axis background/.style={fill=white},
    axis y line*=left,
    axis x line=none,
    tick align=outside,
    xticklabels={F0,Voicing,logHNR,specRastaL1, Loudness,PCM ZCR,Alpharatio},xtick={1,2,3,4,5,6,7},
    ylabel near ticks,
    legend pos=north west,
    legend columns=1,
    legend style={
            /tikz/column 2/.style={
                column sep=2pt,
            }
    }    ] 

\addplot+[mark=o, color=orange,line width=1pt] file {PERTIMIT2.txt};
\node [above=4, text=black] at (axis cs:  1,  16.77) {$16.77$};
\node [above=4, text=black] at (axis cs:  2,  16.99) {$16.99$};
\node [above=8, text=black] at (axis cs:  3,  16.43) {$16.43$};
\node [above=4, left=1, text=black] at (axis cs:  4,  17.46) {$17.46$};
\node [above=3, text=black] at (axis cs:  5,  18.35) {$18.35$};
\node [above=4, text=black] at (axis cs:  6,  17.88) {$17.88$};
\node [above=15, text=black] at (axis cs:  7,  16.46) {$16.46$};

\addlegendentry{PER TIMIT}

\end{axis}

\end{tikzpicture}

\begin{tikzpicture}
\pgfplotsset{set layers}

\begin{axis}[
    width=\linewidth, height=7cm, 
    scale only axis,
    grid = major,
    grid style={dashed, gray!30},
    xmin=0.5,   
    xmax=7.5,  
    ymin=-0.25,   
    ymax=4.5,  
    axis background/.style={fill=white},
    axis y line*=right,
    tick align=outside,
     xticklabels={F0,Voicing,logHNR,specRastaL1, Loudness,PCM ZCR,Alpharatio},xtick={1,2,3,4,5,6,7},
  x tick label style={rotate=60,anchor=east}]
    mark repeat={600},
    ylabel near ticks,
    xlabel near ticks,
    xlabel style={text width=5cm},
    xlabel style={align=center},
    xlabel={test},
    legend pos=north west,
    legend columns=1,
    legend style={
            /tikz/column 2/.style={
                column sep=2pt,
            }
    }    ]
\addplot+[mark=o, color=purple,line width=1pt] file  {labeledcivc.txt};
\node [above=4 , text=black] at (axis cs:  1,  0.02) {$0.02$};
\node [below=5, text=black] at (axis cs:  2,  0.86) {$0.86$};
\node [above=4, text=black] at (axis cs:  3,  0.02) {$0.02$};
\node [below=4, text=black] at (axis cs:  4,  0.77) {$0.77$};
\node [below=3, text=black] at (axis cs:  5,  0.86) {$0.86$};
\node [below=4, text=black] at (axis cs:  6,  0.86) {$0.86$};
\node [above=5, text=black] at (axis cs:  7,  0.06) {$0.06$};

\addlegendentry{CI VC}
\end{axis}
\begin{axis}[
    width=\linewidth, height=7cm, 
    scale only axis,
    grid = major,
    grid style={dashed, gray!30},
    xmin=0.5,   
    xmax=7.5,  
    ymin=7.5,   
    ymax=15,  
    axis background/.style={fill=white},
    axis y line*=left,
    axis x line=none,
    tick align=outside,
    xticklabels={F0,Voicing,logHNR,specRastaL1, Loudness,PCM ZCR,Alpharatio},xtick={1,2,3,4,5,6,7},
    ylabel near ticks,
    legend pos=north west,
    legend columns=1,
    legend style={
            /tikz/column 2/.style={
                column sep=2pt,
            }
    }  ] 
\addplot+[mark=o, color=orange,line width=1pt] file {EERVCreal.txt};
\node [above=4, text=black] at (axis cs:  1,  9.99) {$9.99$};
\node [above=4, text=black] at (axis cs:  2,  9.98) {$9.98$};
\node [above=4, text=black] at (axis cs:  3,  9.08) {$9.08$};
\node [above=4, left=1, text=black] at (axis cs:  4,  9.32) {$9.32$};
\node [above=3, text=black] at (axis cs:  5,  12.68) {$12.68$};
\node [above=4, text=black] at (axis cs:  6,  10.1) {$10.1$};
\node [above=4, text=black] at (axis cs:  7,  10.01) {$10.01$};

\addlegendentry{EER VC}

\end{axis}

\end{tikzpicture}

\end{adjustbox}  
\caption {Left : Phone Error Rate and CI estimate values on TIMIT for every considered pseudo-label --- Right: Equal Error Rate and CI estimate values on VoxCeleb for every considered pseudo-label. Error rates appear on the left y axis. We can observe the monotonic relation between the estimator and the downstream errors, particularly for TIMIT.}
\label{results}
\end{figure}
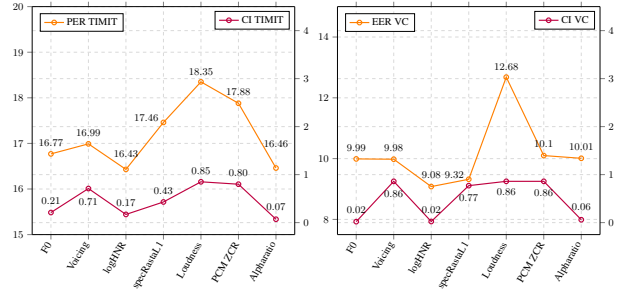

Figure \ref{results} summarizes the results of the experiment for all the considered pseudo-labels, reporting the CI estimates and the downstream performance for each of the two tasks. It shows the evolution of the conditional independence estimator and the PER and EER, respectively on TIMIT and VoxCeleb. Despite a little bump on the \textit{loudness} pretraining, the two curves seem to follow the same trajectories.

We are looking for a monotonic relationship between CI estimates and the downstream error. Two classic assessors of monotony are considered: Spearman Correlation and Kendall Tau. When Pearson correlation measures the linear correlation between the values, Spearman correlation is a Pearson Correlation on the ranks of the values. Kendall $\tau$ considers all the pairs of pseudo-labels, and checks whether their order in the CI estimate is the same for the error rate (\ie the pair is concordant ). The more concordant pairs there are, the higher Kendall $\tau$ is.

Spearman correlations reach \textbf{0.48} for speaker recognition and a high \textbf{0.93} on TIMIT for ASR, while Kendall $\tau$ is respectively \textbf{0.41} and \textbf{0.81} for the two tasks. The correlations between CI and the downstream error are logically positive. As the lower the CI estimate is, the more independent is the pseudo-label from the speech samples given the label, the lower is the downstream error, confirming theoretical insights\cite{lee2020predicting}.
Finally, to test the influence of the downsampling method on our estimate, we compute the HSIC values based on vectors downsampled with SVCCA \cite{raghu2017svcca}. It led to minor differences with a mean relative difference of $1.5$\% on the final CI estimates. This hints to the robustness of our method to downsampling method variation.

\section{Combining Pseudo-labels}
Finding the best combination of pseudo-labels certainly involves more than individual estimates as may intervene questions of shared information. Nevertheless, we wanted to test our estimator with pseudo-labels regrouped in a naive way. In a second experiment, for each task, two self-supervised models are trained to predict  two different groups of pseudo-labels. One learns to predict jointly the ones with the best CI estimator scores, and one learns the worst pseudo-labels according to our estimator.\cmmnt{In both cases, we also learn to reconstruct Mel spectrograms and to produce MFCCs.} The same experimental setup is kept with one slight change: in this experiment, one of the objectives was to push further the results. So, the encoder parameters were not frozen but were updated during the retraining, with an SGD optimizer.

Pseudo-labels selected and results are described in Table \ref{tab:regrouping}. The third column shows EER for the VoxCeleb (VC) experiments, and PER for the TIMIT (TIM) ones. As expected, the results obtained with the best pseudo-labels are better than the ones with the worst ones. Besides that, results obtained with the non-freezed features are better than with freezed ones. This is probably due to the big distributional shift from the pretraining dataset (Common Voice) and the downstream ones. Unfreezing the encoder parameters may allow the encoder to adapt to the new points' distribution.

\section{Conclusion}
In this work, we introduce an estimator of the utility of a given pretext task as a function of the downstream task to better explain and motivate the selection of pretext tasks in self-supervised learning settings. The estimator evaluates the conditional independence between the pretext label and the speech samples given the downstream labels, using HSIC as the independence criterion. The conducted experiments validate the proposed utility estimator on two tasks: ASR and speaker recognition. This opens a range of possibilities for finding and selecting new pretext tasks in self-supervised learning for speech or other types of data.

\section{Acknowledgements}
We want to thank Zoltan Szabo for the discussions we had on conditional independence and thank as well all the SpeechBrain library contributors. This work is partly funded by L’Agence de l’innovation de défense.
\clearpage

\bibliographystyle{IEEEtran}

\bibliography{mybib}


\end{document}